\documentclass[prl,twocolumn,aps,superscriptaddress,letterpaper]{revtex4-1}
\usepackage{amsmath,amssymb}
\usepackage{epsfig}
\usepackage{graphicx}
\usepackage{amsmath}
\usepackage{amsfonts}
\usepackage{epstopdf}
\usepackage{bbold}
\usepackage{color}
\usepackage{microtype}
\usepackage{dsfont}  
\usepackage{slashed}

\newcommand{\be}{\begin{equation}}
\newcommand{\ee}{\end{equation}}
\long\def\bea#1\eea{\be\begin{split} #1 \end{split}\ee}
\newcommand{\eq}[1]{Eq.~(\ref{eq:#1})}
\newcommand{\figu}[1]{Fig.~\ref{fig:#1}}

\newcommand{\nn}{\nonumber}
\newcommand{\ba}{\begin{array}}
\newcommand{\ea}{\end{array}}

\newcommand{\R}{{\mathds R}}
\newcommand{\C}{{\mathds C}}
\newcommand{\re}{\mathop{\rm{Re}}}
\newcommand{\im}{\mathop{\rm{Im}}}

\def\av#1{ \left\langle #1 \right\rangle }

\newcommand{\Man}{\mathcal{M}}

\usepackage[colorlinks=true,backref=false, linktocpage=true,
citecolor=blue,urlcolor=blue,linkcolor=blue,pdfpagemode=UseOutlines]{hyperref}

\hypersetup{%
  bookmarksnumbered=true,
  pdftitle = {},
  pdfsubject = {},
  pdfauthor = {},
  pdfkeywords = {}
}

\def\Tr {\mathop{\hbox{Tr}}}

\begin{document}

\title{Monte Carlo calculations of the finite density Thirring model}

\author{Andrei Alexandru}
\email{aalexan@gwu.edu }
\affiliation{Department of Physics, The George Washington University,
Washington, DC 20052}
\affiliation{Department of Physics,
University of Maryland, College Park, MD 20742}
\author{G\"ok\c ce Ba\c sar}
\email{gbasar@umd.edu}
\author{Paulo F. Bedaque}
\email{bedaque@umd.edu}
\author{Gregory W. Ridgway}
\email{gregridgway@gmail.com}
\author{Neill C. Warrington}
\email{ncwarrin@umd.edu}
\affiliation{Department of Physics,
University of Maryland, College Park, MD 20742}

\date{\today}

\begin{abstract}
We present results of the numerical simulation of the two-dimensional Thirring model at finite density and temperature. The severe sign problem is dealt with by  deforming the domain of integration into complex field space. This is the first example where a fermionic sign problem is solved in a quantum field theory by using the holomorphic gradient flow approach, a generalization of the Lefschetz thimble method. 
\end{abstract}


\maketitle


\section{Introduction}
Monte Carlo calculations are frequently the only approach available to study certain strongly interacting systems. Despite great progress in many areas of both physics and chemistry, the use of Monte Carlo methods is limited to problems that can be formulated in imaginary (as opposed to real) time and in the absence of chemical potentials. This limitation excludes a vast array of interesting transport and non-equilibrium observables as well as the equilibrium properties of systems with a finite density of a conserved charge. Dense strongly interacting matter, a system of major concern in nuclear physics, is one such excluded case, along with many important cases in condensed matter, such as strongly correlated electronic systems. The reason for this limitation is 
that observables are obtained by averaging contributions with different complex phases which nearly cancel out. This is the famous ``sign problem".

A new idea to solve the sign problem was put forward in~\cite{Cristoforetti:2012su}. It consists in complexifying the fields (the variables in the path integral) and changing the functional integration region  to a certain manifold embedded in the space of these complex variables.
Originally the multidimensional analogue of the stationary-phase contour, the ``Lefschetz thimbles", was suggested as an optimal choice of integration manifold. However, our method uses  manifolds interpolating between the real hyperplane and the Lefschtez thimbles. These interpolating manifolds have numerous computational advantages over the thimbles 
for a variety of reasons, which we discuss later. 
 The original idea sparked a flurry of interest leading to 
algorithmic development~\cite{Cristoforetti:2014gsa,Fukushima:2015qza,Alexandru:2016lsn,Alexandru:2015sua}
and subsequent applications in many simple models including bosonic theories~\cite{Fujii:2013sra,Mukherjee:2013aga,Cristoforetti:2013wha,Cristoforetti:2013qaa,Tanizaki:2014tua,DiRenzo:2015foa,Alexandru:2016san,Alexandru:2016lsn}, 
fermionic toy models~\cite{Mukherjee:2014hsa,Fujii:2015bua,Tanizaki:2015rda,Fujii:2015vha,Fujii:2015rdd,Alexandru:2015xva,Alexandru:2015sua} where the sign problem is usually more difficult to solve, and in even real time dynamics~\cite{Alexandru:2016gsd}.
The purpose of the present paper is to describe the first calculation of this type in an 
interacting fermionic field theory which shares common properties with QCD.

\section{Thirring model}

The model we study in this paper is defined in the continuum by the Euclidean action
\be\label{eq:S_continuum}
S=\int d^2x\ [\bar\psi^\alpha (\slashed{\partial}+\mu \gamma_0 +m)\psi^\alpha 
+ \frac{g^2}{2N_F}\bar\psi^\alpha\gamma_\mu\psi^\alpha \bar\psi^\beta\gamma_\mu\psi^\beta],
\ee 
where the flavor indices take values $\alpha,\beta=1,\ldots,N_F$, $\mu$ is the chemical potential and the Dirac spinors $\bar\psi,\psi$ have two components. It is convenient to treat the four-fermion interaction by introducting an auxiliary vector field $A_\mu$. The path integration over $A_\mu$ of the action: 
\be\label{eq:S_continuum_aux}
\!\!\!S=\int d^2x\ \left[ \frac{N_F}{2g^2}A_\mu A_\mu
+\bar\psi^\alpha (\slashed{\partial}+\mu \gamma_0 +i \slashed{A}+m)\psi^\alpha \right]
\ee  
generates~\eq{S_continuum}.  We use two discretizations of \eq{S_continuum_aux}. The Wilson lattice action is given by
\be
\!\!\!S=\sum_{x,\nu} \frac{N_F}{g^2} (1-\cos A_\nu(x))+\sum_{x,y} \bar\psi^\alpha(x) D^{W}_{xy}(A)  \psi^\alpha(y).
\ee 
with
\begin{eqnarray}
D^W_{xy} = &\delta_{xy} - \kappa \sum_{\nu=0,1}  
\Big[ 
 (1-\gamma_\nu) e^{i A_\nu(x)+\mu \delta_{\nu 0}} \delta_{x+\nu, y}
   \nonumber \\
& + (1+\gamma_\nu) e^{-i A_\nu(x)-\mu \delta_{\nu 0}}  \delta_{x, y+\nu}
\Big], 
\end{eqnarray}
and $\kappa=1/(2m+4)$. For even $N_F$ we can also use
the staggered (Kogut-Susskind) lattice action:
\be
\!\!\!\!\!S=\sum_{x,\nu} \frac{N_F}{g^2} (1-\cos A_\nu(x)) 
+
\sum_{x,y} \bar\chi^\alpha(x) D^{KS}_{xy}(A)  \chi^\alpha(y)\,,
\ee 
with 
\bea
D_{xy}^{KS} = m +\frac{1}{2}\sum_{\nu=0,1}
&\Big[ 
 \eta_\nu e^{i A_\nu(x)+\mu \delta_{\nu 0}} \delta_{x+\nu, y}  
 \\
&-\eta_\nu^\dagger  e^{-i A_\nu(x)-\mu \delta_{\nu 0}}  \delta_{x, y+\nu} \Big]\,.  
\eea 
Here $\alpha=1,\cdots, N_F/2$,  $\bar\chi, \chi$ are Grassmann numbers with no spinor indices and $\eta_0(x) =1, \eta_1= (-1)^{x_0}$. In either discretization, the integration over the fermion fields leads to
\be
\!\!\!S=N_F 
\left(  
\frac{1}{g^2}\sum_{x,\nu} (1-\cos A_\nu(x)) -\gamma \log\det D(A)
\right),
\ee with $\gamma=1$ (Wilson), or $\gamma=1/2$ (staggered). 
Both of these lattice actions  describe $N_F$ Dirac fermions in the continuum. For $\mu\neq 0$ the determinant $\det D(A)$ is not real so this model cannot be simulated by standard Monte Carlo techniques. In this work we use $N_F=2$.

\begin{figure*}[t]
\includegraphics[scale=0.85]{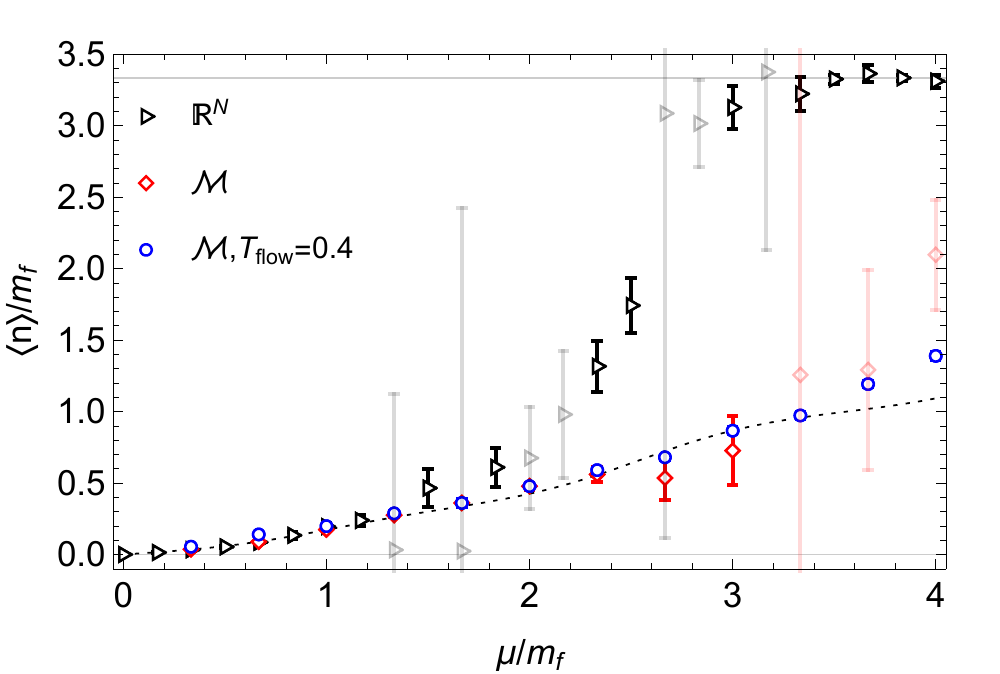}
\includegraphics[scale=0.85]{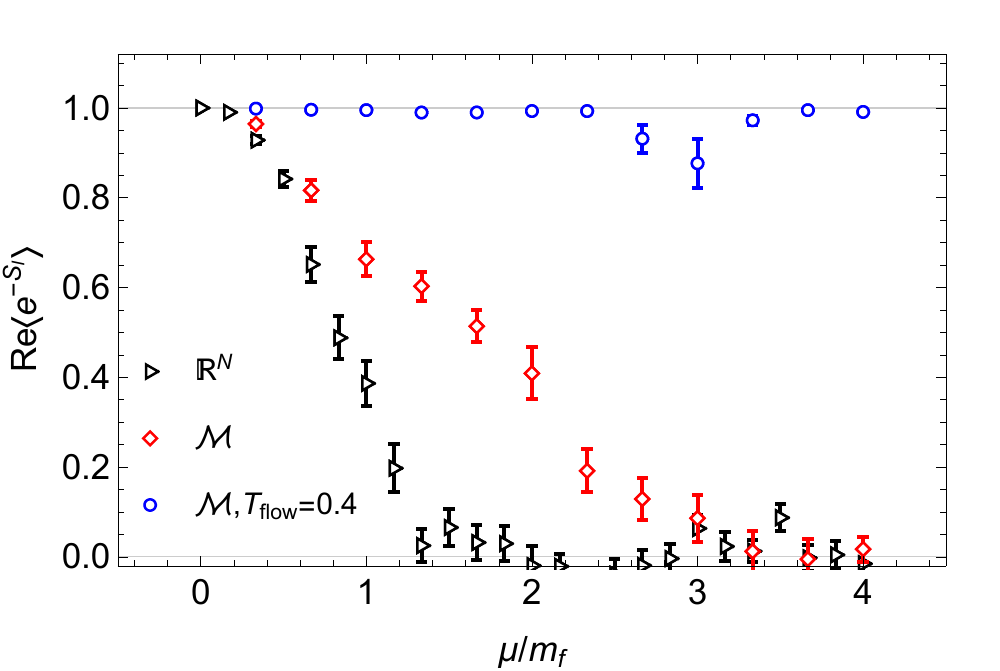}
\caption{Fermion density per flavor as a function of the chemical potential $\mu$ (left) and average sign (right). In the left plot, 
the upper horizontal line is the saturation density and the dotted curve corresponds to the free gas result; we grayed out the points 
with the errorbars exceeding $0.3$ to make the figure easier to read.}
\label{fig:sign_problem}
\end{figure*}

\section{the algorithm}


Here we summarize the algorithm we use  and the mathematical results associated with it~\cite{Alexandru:2015xva,Alexandru:2015sua}. 
The main idea is to deform the domain of integration in field space where the path integral is performed ($\R^N$), justified by the Cauchy's theorem, into a submanifold $\Man$ of complex space ($\C^N\approx \R^{2N}$) in such a way as to ameliorate the sign problem:
\be
\langle \mathcal{O}\rangle =
\frac{\int_{\R^N} d\phi_i \  e^{-S[\phi]} \mathcal{O}[\phi] }{\int_{\R^N}  d\phi_i \  e^{-S[\phi]}}
=
\frac{\int_{\Man} d\phi_i \  e^{-S[\phi]} \mathcal{O}[\phi] }{\int_{\Man}  d\phi_i \  e^{-S[\phi]}},
\ee where $\phi_i$, $i=1,\ldots, N$ are real variables but $S[\phi]$ is {\it not} real. The integral over $\Man$
can be written by using a parametrization $\phi(\zeta)$ in terms of the real parameters $\zeta$:
\be
Z = \int_{\Man} d\phi_i \ e^{-S[\phi] }
= \int_{\R^N} d\zeta_i\      {\rm det}\left ( \frac{\partial \phi_i}{\partial \zeta_j}\right )  e^{-S[\phi(\zeta)]}.
\ee 
How can $\Man$ be chosen so the sign problem is improved? One answer is to consider the manifold obtained by taking  every point in the original integration domain ($\R^N$) as an initial condition and evolving it according to the {\em holomorphic gradient flow} equations
\bea\label{eq:flow}
\frac{d\phi_i}{dt} = \overline{ \frac{\partial S}{\partial\phi_i}}, \quad \phi_i(0) = \zeta_i, 
\eea by a fixed ``time" $T$.
The transport of an orthonormal basis in $\R^N$ with the flow is determined by the matrix $J_{ij}(T)$  that satisfies
\bea\label{eq:flowJ}
\frac{dJ_{ij}}{dt} = \overline{H_{ij}}\overline{ J_{kj}}, \quad H_{ij}\equiv \frac{\partial ^2S}{\partial\phi_i\partial\phi_k},\quad J_{ij}(0) = \delta_{ij},
\eea 
with $\det J =  {\rm det} \frac{\partial \phi_i}{\partial \zeta_j}$ being the Jacobian.

The imaginary part of the action $S_I$ is constant along the flow lines of \eq{flow} while the real part $S_R$ grows monotonically.
We will now argue that $\Man$, defined by flowing $\R^N$ a fixed amount is an allowed choice of domain of integration.
Assuming the integrand $\text{e}^{-S}$ has no singularities at finite values of $\phi_i$---as is the case for all field theories---the only obstacle to the deformation of integration domain can occur when the fields approach infinity, where singularities typically appear. Thus, unless at some intermediate stage we encounter a singularity at infinity, the integral remains unchanged under the deformation~\footnote{A clear discussion of the classification of all the possible integration domains and related topics in a physicist's language is found in \cite{Witten:2010zr}; more mathematically oriented discussions can be found in~\cite{fedoryuk,pham,kaminski}.}. On the other hand, the flow in~\eq{flow}, increases $S_R$ and consequently decreases the absolute value of the integrand $|e^{-S}|=e^{-S_R}$. Therefore, starting from a convergent integral over $\R^N$\footnote{The path integral {\it at finite lattice spacing} must be convergent. Standard renormalization procedure is required to define the continuum limit.}, and deforming the domain by the flow, we never encounter a divergence at infinity and the integral over $\Man$ is equal to the integral over $\R^N$. The choice of $\Man$ as the integration manifold is not only legitimate but also profitable in taming the sign problem.
In fact, notice that for large $T$ the flow pins certain points in $\R^N$ to the critical points satisfying $\partial S/\partial\phi_i=0$ as the flow cannot continue past it. The infinitesimal neighborhood around each of these points flows to an $N$ dimensional manifold attached to the critical point, called a ``Lefschetz thimble'' (multi-dimensional stationary phase contour). Points along other directions flow to regions with large $S_R$. The flow is tangent to the thimbles and, as such, cannot cross them. As $T\rightarrow \infty$, $\Man$ asymptotically approaches  the particular combination of thimbles  equivalent to the original path integral. $S_I$ is constant on each thimble and for that reason it was advocated in the past as the best domain to deform to in order solve the sign problem~\cite{Cristoforetti:2012su}. However, thimbles are separated by large action barriers, making it difficult to tunnel to all relevant thimbles in a Monte Carlo. Instead, by varying $T$ it is possible to generate alternative manifolds that interpolate between $\R^N$ and the sum over thimbles (where $S_I$ is piecewise constant). The amount of flow controls simultaneously the severity of the sign problem and the depth of the action barriers. 

For fermionic systems the zeros of the determinant form boundaries for thimbles, some of these bounding multiple thimbles.
The integrand remains a holomorphic function since the determinant is a polynomial in the field variables. However, the 
action has logarithmic singularities and its gradient has poles at these points. These singularities 
attract the flow and a subset of the configuration space flows into these points in finite flow time, including some of the 
points on the parametrization manifold. It is then the case that the flowed manifold includes a set of determinant zeros, 
often forming cusps at these points. 
However, since the flow always moves in the direction of increasing $S_R$, when it flows into
these singularities, it approaches them from directions where $e^{-S_R}$ monotonically decreases. Consequently, in simulations, proposals that flow into these points are rejected since they have infinite action, and consequently zero acceptance probability.

The algorithm we use is the  Metropolis algorithm applied to the variables $\zeta_i$ and the effective action $S_\text{eff}[\zeta] =  S[\phi(\zeta)] - \log\det J$. The configurations are sampled according to $\re S_\text{eff}[\zeta] = S_R[\phi(\zeta)]-\log|\det J|$ and
the phase $\varphi(\zeta)\equiv\im S_\text{eff}[\zeta]=  S_I[\phi(\zeta)]-{\rm arg} \det J$ is included through reweighting according to the relation
\bea
\av{ \mathcal{O}} &=
\frac{
\int d\zeta_i\   \mathcal{O} \det J e^{-S[\phi(\zeta)]}
}
{
\int d\zeta_i\    \det J e^{-S[\phi(\zeta)]}
}\nn\\
&=
\frac{
\int d\zeta_i\   \mathcal{O} e^{-i \varphi(\zeta)}  e^{-\re S_\text{eff}[\zeta]}
}
{
\int d\zeta_i\   e^{-\re S_\text{eff}[\zeta]}
}
\frac{
\int d\zeta_i\   e^{-\re S_\text{eff}[\zeta]}
}
{
\int d\zeta_i\    e^{-i \varphi(\zeta)} e^{-\re S_\text{eff}[\zeta]}
}\\
&=
\frac
{\av{ \mathcal{O} e^{-i \varphi(\zeta)} }_{\re S_\text{eff}}}
{\av{ e^{-i \varphi(\zeta)} }_{\re S_\text{eff}}}.
\eea 
The integration domain for all integrals above is $\R^N$.
In all cases we have explored the {\em residual phase}, $e^{i\im J}$, varies slowly. 
The phase $e^{-i S_I}$ is highly oscillating on $\R^N$ but its fluctuations are reduced as the flow time increases (see right panel of~\figu{sign_problem}.)

Either action, Wilson or staggered, has a critical point at $A_0(x)=i A, A_1(x)=0$, constant in spacetime, satisfying:
\be
i\sinh A = \gamma g^2 \Tr \frac{\partial D}{\partial A_0(x)} D^{-1} \,.
\ee 
The tangent space to the thimble at this point is purely real and is obtained by a simple translation of 
$\R^N$ through $A_0(x) \rightarrow A_0(x)+i A, A_1(x)\rightarrow A_1(x)$, which we name as the ``main tangent space". 
Since the action is periodic in each of the variable $A_\mu(x)$ (with period $2\pi$) we can shift the integration manifold by a constant in the imaginary direction without introducing any singularities, ensuring the integral remains unchanged. 
As such the main tangent space is a legitimate manifold over which to perform the path integral. It is also 
an approximation of a thimble. Thus, as we will see below, in some cases shifting to the main tangent space is sufficient to circumvent the sign problem.


To sample efficiently the configurations $\zeta$ in the main tangent space $\cal M$, 
we make proposals that take into account the effect of the flow map $\zeta\to \phi(\zeta)$, which
contracts and expands various directions in tangent space at different rates. We use
the ``eigenvalues'' and ``eigenvectors'' of the Hessian at the critical point $H_0$: $H_0 v_i = \lambda_i \overline{v_i}$.
The ``eigenvectors'' $v_i$ corresponding to positive eigenvalues span $\cal M$. A shift in direction
$v_i$ is proposed with magnitude $\epsilon/\sqrt{\lambda_i} \exp(-T\lambda_i)$, with $\epsilon$
a random variable uniformly distributed over the interval $[-\Delta,\Delta]$. $\Delta$ is tuned to get
a good acceptance rate~\cite{Alexandru:2015xva}. To reduce the computational cost, we used an estimator 
for $\det J$ introduced in \cite{Alexandru:2016lsn}. 

\section{Results}

To determine the physical parameters of the discretized theory, we measure two particle masses: a fermion and a boson.
Denoting the lattice spacing with $a$, 
the dimensionless masses $a m_f$ and $a m_b$ are determined by fitting the large time behavior of the correlators 
$\av{{\cal O_{\alpha}}(t){\cal O_{\alpha}}(0)^\dagger}$ with an exponential $\exp[-(am_{\alpha}) (t/a)]$ with
${\cal O}_f = \psi_1$ and ${\cal O}_b=\bar\psi_i \gamma_5 (\tau_3)_{ij}\psi_j$, 
where the subscripts indicate the flavors.
For the free theory ($g=0$) we have $m_b=2m_f$, and the ratio $m_b/m_f$ can be used to gauge the strength of the interaction. When $m_b/m_f\ll2$, 
the theory is strongly interacting.

As an illustration of our method we consider the results obtained with the Wilson action in a $10\times 10$ lattice and parameters $g=1.0$ and $m=-0.250$. For these parameters we find that the fermion has a mass of $a m_f=0.30(1)$ and the boson mass is $a m_b=0.44(1)$, showing that these parameters correspond to a  strongly coupled theory.
In \figu{sign_problem} we show the average fermion density (per flavor) $\langle n\rangle $ on the left  and the average sign $\langle  e^{-i S_I}\rangle$ on the right. The results obtained by standard reweighting on $\R^N$ are shown in black. It is clear that as soon as $\mu \approx m_f$ the average sign drops to zero and reweighting leads to large uncertainties. This is the basic manifestation of the sign problem. In red we show the results obtained by an integration over the main tangent plane (which is no more computationally expensive than an integration over $\mathds{R}^N$). The average phase  approaches zero at a much larger value of $\mu$ and the error bars in $\langle n\rangle $ reflect that. So it's possible to peer deeper into the phase diagram by simply shifting the domain of integration into complex space. Tangent plane calculations do not allow for calculations above $\mu \simeq 2.5 m_f$ but the remaining sign problem can be cured by using a manifold $\Man$ obtained from $\R^N$ by flowing by a ``time" $T=0.4$. The results of this calculations are shown in blue.  In \figu{sign_problem} we also include the result of a free fermion gas with mass equal to $m_f$. The agreement between the free theory calculation and this interacting model is expected, for at these low temperatures the equilibrium state contains mostly particles (as opposed to anti-particles) and particles interact weakly among themselves at low momenta due to the Pauli principle.

\begin{figure}[t]
\includegraphics[width=\columnwidth]{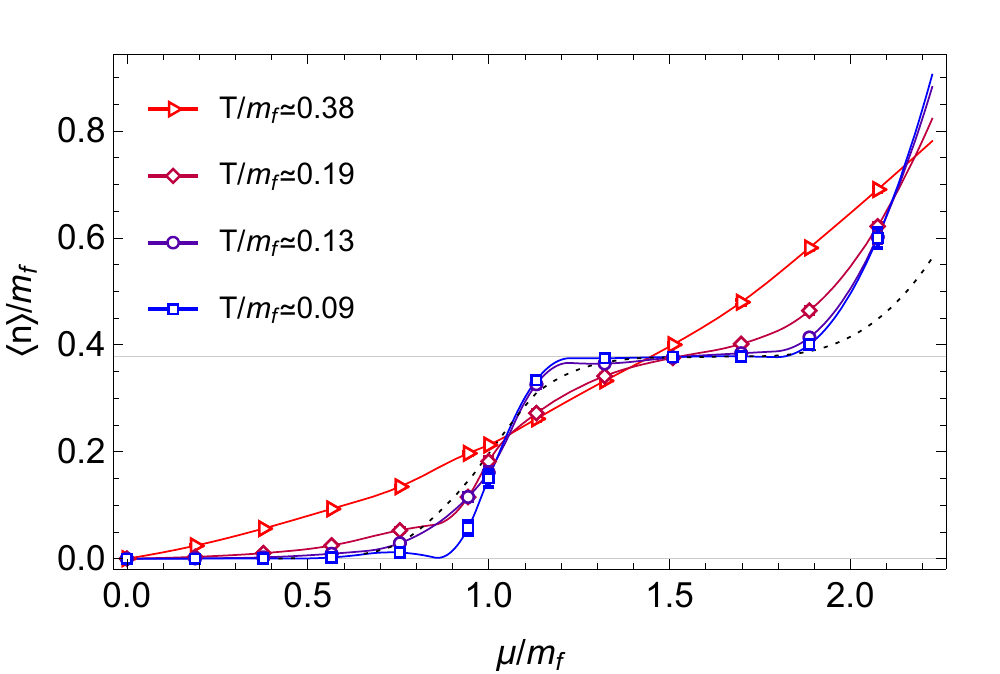}
\caption{$\langle n \rangle$ as a function of $\mu$ for several temperatures. The horizontal line is the density 
that corresponds to one particle in the box (per flavor.)
The solid curves are splines interpolations of the
data points to guide the eye. The dotted curve represents a free fermion gas in the staggered discretization 
on a $40\times10$ lattice with the mass set to the value of $a m_f = 0.265$.}
\label{fig:blaze}
\end{figure}

In \figu{blaze}  we extend the previous results to lower temperatures and demonstrate that our method can handle temperatures exhibiting the ``Silver Blaze"  phenomenon~\cite{Cohen:2003kd}, that is, the independence  of observables to the value of $\mu$ below a threshold value (of the order of the lightest fermion). Our results clearly show the plateaus associated to the Silver Blaze phenomenon. The first threshold is near $\mu \approx m_f$. \footnote{It is also possible to obtain information on the two-particle interactions by analyzing the second plateau as discussed in \cite{Bruckmann:2015hua}, which we left for future work.} This result is not trivial for two reasons. First, other methods dealing with the sign problem have difficulties dealing with Silver Blaze situations \cite{Hayata:2015lzj}. Second, there is a worry that our sampling can become trapped near a  local minima of $S_R$ (corresponding to a single thimble) at the exclusion of other important minima. Greater flow makes the landscape of Boltzmann weights $e^{-S_R}$ more mountainous, isolating local minima from each other, which potentially causes a problematic situation for an algorithm based on a Monte Carlo chain.
As pointed out in~\cite{Tanizaki:2015rda}, a defective sampling of field space that erroneously samples  only the main thimble washes out the staircase structure in the $\langle n \rangle $ vs. $\mu$ plot in favor of a straight line. Thus, seeing the staircase in \figu{blaze} is strong evidence that our sampling is sufficiently ergodic.

\begin{figure}[t]
\includegraphics[width=\columnwidth]{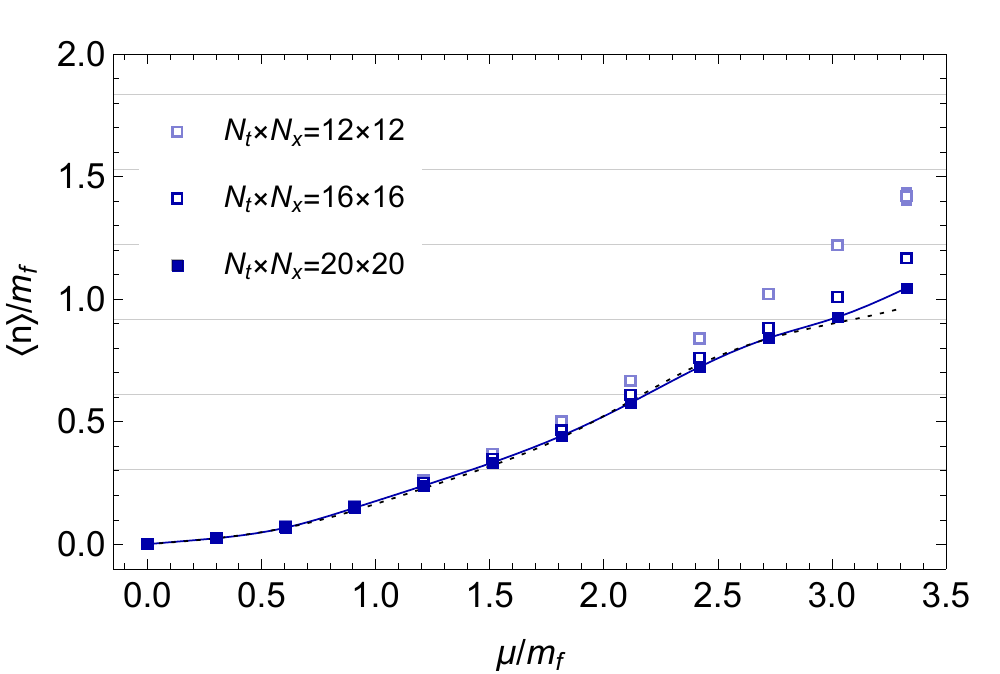}
\caption{Particle density as a function of the chemical potential for different lattice spacings, for fixed volume ($m_f L\approx3.31$) 
and temperature ($T/m_f\approx0.302$). 
The solid line represents a spline interpolating through finest lattice spacing data points. The dotted line represents the 
fermion free gas result. Horizontal lines indicate integral number of particles in the box.}
\label{fig:whole_enchillada}
\end{figure}

We find that the severity of the sign problem varies little as the lattice spacing is varied at fixed volume.
For the staggered action we carried out three sets of simulations at different lattice spacing,
on lattices sizes $20\times 20, 16\times 16$ and $12\times 12$.
The parameters were tuned such that all physical observables---the temperature, the volume and the fermion and boson mass---were the 
same in physical units. The quantity $a m_f$ was tuned to be in the ratio $\frac{20}{20}: \frac{20}{16} : \frac{20}{12}$. For these simulations $m_b/m_f\approx 1.70$. 
The results are summarized in \figu{whole_enchillada}.
  
\begin{figure}[h]
\includegraphics[width=\columnwidth]{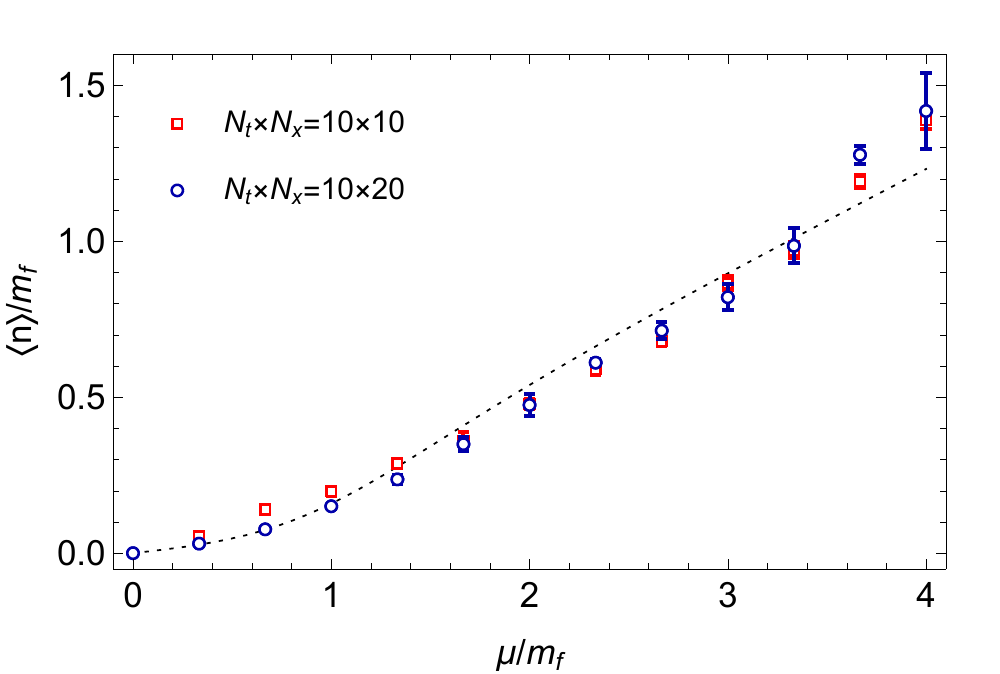}
\caption{Fermion density as a function of chemical potential on two different volumes, $10\times10$ from \figu{sign_problem}
and $10\times20$ using the same parameters. Dotted line is the free gas result.}
\label{fig:thermolimit}
\end{figure}

To assess proximity to the thermodynamic limit, we compare the results included in \figu{sign_problem} with the results obtained from a system at equal temperature, but with twice the spatial extent. The results are presented in \figu{thermolimit}. We find that the density varies little as the spatial extent of the system is doubled, indicating that we are close to the thermodynamic limit. 

 There are some general trends in the scaling of the computational cost with the degrees of freedom. The continuum limit does not pose any particular challenge besides the fact that the evaluation of $\det J$ and the fermion determinant has a computational cost proportional to $N^3 \sim V^3$ ($V$ is the spacetime volume). 
 On the other hand, both the increase of the physical volume and the lowering of the temperature requires more work. The sign problem becomes more severe and we need to use a large flow time to cure it. 

\section{Discussions and prospects}

We have solved the sign problem of the finite density Thirring model by deforming the domain of integration 
of the path integral into complex space. For some regions of the parameter space, a simple shift of fields suffices to tame the sign problem. In other regions the holomorphic flow is required. The method we use has the advantage that 
it does not
require an explicit thimble decomposition, which is a highly non-trivial problem for quantum field theories. For fermionic
theories the zeros of the determinant play an important role in the decomposition and, in principle, could interfere with the holomorphic flow. We do not see any evidence of such problems and our results are in excellent agreement with theoretical expectations.
The method is general and should be applicable to other theories of physical interest. 


%

\section{Acknowledgments }
A.A. is supported in part by the National Science Foundation CAREER grant PHY-1151648 and 
the U.S. Department of Energy grant DE-FG02-95ER40907. A.A. gratefully acknowledges the hospitality 
of the Physics Departments at the University of Maryland and the University of Kentucky where part of this work was carried out.
G.B., P.F.B., G.R. and N.C.W. are supported by U.S. Department of Energy under Contract No. DE-FG02-93ER-40762.

\bibliography{thimbles}
\end{document}